\documentclass[aps,amsmath,amssymb, prc,twocolumn,superscriptaddress]{revtex4-1}

\usepackage{graphicx}
\usepackage{dcolumn}
\usepackage{bm}
\usepackage[utf8]{inputenc}
\usepackage[ngerman,english]{babel}
\usepackage{float}
\usepackage{multirow}
\usepackage{longtable}
\usepackage{dcolumn}
\newcolumntype{d}{D{.}{.}{-1}}
\usepackage{setspace}
\usepackage{threeparttable}
\usepackage{tabularx}
\usepackage[symbol*]{footmisc}
\DefineFNsymbolsTM{myfnsymbols}{
  \textasteriskcentered *
  \textdagger    \dagger
  \textdaggerdbl \ddagger
  \textsection   \mathsection
  \textbardbl    \|%
  \textparagraph \mathparagraph
}%
\setfnsymbol{myfnsymbols}

\usepackage{siunitx}

\begin{document}


\title{Experimental identification of the $T = 1, J^{\pi} = 6^+$ state of $^{54}$Co and isospin symmetry in $A = 54$ studied via one-nucleon knockout reactions}


\author{M. Spieker}
\altaffiliation{Present address: Department of Physics, Florida State University, Tallahassee, Florida 32306, USA}
\affiliation{National Superconducting Cyclotron Laboratory, Michigan State University, East Lansing, Michigan 48824, USA}

\author{D. Weisshaar}
\affiliation{National Superconducting Cyclotron Laboratory, Michigan State University, East Lansing, Michigan 48824, USA}

\author{A. Gade}
\affiliation{National Superconducting Cyclotron Laboratory, Michigan State University, East Lansing, Michigan 48824, USA}
\affiliation{Department of Physics and Astronomy, Michigan State University, East Lansing, Michigan 48824, USA}

\author{B.\ A. Brown}
\affiliation{National Superconducting Cyclotron Laboratory, Michigan State University, East Lansing, Michigan 48824, USA}
\affiliation{Department of Physics and Astronomy, Michigan State University, East Lansing, Michigan 48824, USA}

\author{P. Adrich}
\altaffiliation{Present address: National Centre for Nuclear Research, Otwock, Poland}
\affiliation{National Superconducting Cyclotron Laboratory, Michigan State University, East Lansing, Michigan 48824, USA}

\author{D. Bazin}
\affiliation{National Superconducting Cyclotron Laboratory, Michigan State University, East Lansing, Michigan 48824, USA}
\affiliation{Department of Physics and Astronomy, Michigan State University, East Lansing, Michigan 48824, USA}

\author{M.\ A. Bentley}
\affiliation{Department of Physics, University of York, Heslington, York YO10 5DD, United Kingdom}

\author{J.\ R. Brown}
\affiliation{Department of Physics, University of York, Heslington, York YO10 5DD, United Kingdom}

\author{C.\ M. Campbell}
\altaffiliation{Present address: Nuclear Science Division, Lawrence Berkeley National Laboratory, Berkeley, California 94720, USA}
\affiliation{National Superconducting Cyclotron Laboratory, Michigan State University, East Lansing, Michigan 48824, USA}
\affiliation{Department of Physics and Astronomy, Michigan State University, East Lansing, Michigan 48824, USA}

\author{C.\ Aa. Diget}
\affiliation{Department of Physics, University of York, Heslington, York YO10 5DD, United Kingdom}

\author{B. Elman}
\affiliation{National Superconducting Cyclotron Laboratory, Michigan State University, East Lansing, Michigan 48824, USA}
\affiliation{Department of Physics and Astronomy, Michigan State University, East Lansing, Michigan 48824, USA}

\author{T. Glasmacher}
\affiliation{National Superconducting Cyclotron Laboratory, Michigan State University, East Lansing, Michigan 48824, USA}
\affiliation{Department of Physics and Astronomy, Michigan State University, East Lansing, Michigan 48824, USA}

\author{M. Hill}
\affiliation{National Superconducting Cyclotron Laboratory, Michigan State University, East Lansing, Michigan 48824, USA}
\affiliation{Department of Physics and Astronomy, Michigan State University, East Lansing, Michigan 48824, USA}

\author{B. Longfellow}
\affiliation{National Superconducting Cyclotron Laboratory, Michigan State University, East Lansing, Michigan 48824, USA}
\affiliation{Department of Physics and Astronomy, Michigan State University, East Lansing, Michigan 48824, USA}

\author{B. Pritychenko}
\affiliation{National Nuclear Data Center, Brookhaven National Laboratory, Upton, New York 11973, USA}

\author{A. Ratkiewicz}
\altaffiliation{Present address: Lawrence Livermore National Laboratory, Livermore, California, 94550, USA}
\affiliation{National Superconducting Cyclotron Laboratory, Michigan State University, East Lansing, Michigan 48824, USA}
\affiliation{Department of Physics and Astronomy, Michigan State University, East Lansing, Michigan 48824, USA}

\author{D. Rhodes}
\affiliation{National Superconducting Cyclotron Laboratory, Michigan State University, East Lansing, Michigan 48824, USA}
\affiliation{Department of Physics and Astronomy, Michigan State University, East Lansing, Michigan 48824, USA}

\author{J.\ A. Tostevin}
\affiliation{Department of Physics, Faculty of Engineering and Physical Sciences, University of Surrey, Guildford, Surrey GU2 7XH, United Kingdom}


\date{\today}

\begin{abstract}

New experimental data obtained from $\gamma$-ray tagged one-neutron and one-proton knockout from $^{55}$Co is presented. A candidate for the sought-after $T=1, T_z = 0, J^{\pi} = 6^+$ state in $^{54}$Co is proposed based on a comparison to the new data on $^{54}$Fe, the corresponding observables predicted by large-scale-shell-model (LSSM) calculations in the full $fp$-model space employing charge-dependent contributions, and isospin-symmetry arguments. Furthermore, possible isospin-symmetry breaking in the $A=54$, $T=1$ triplet is studied by calculating the experimental $c$ coefficients of the isobaric mass multiplet equation (IMME) up to the maximum possible spin $J=6$ expected for the $(1f_{7/2})^{-2}$ two-hole configuration relative to the doubly-magic nucleus $^{56}$Ni. The experimental quantities are compared to the theoretically predicted $c$ coefficients from LSSM calculations using two-body matrix elements obtained from a realistic chiral effective field theory potential at next-to-next-to-next-to-leading order (N$^3$LO).

\end{abstract}


\maketitle


The atomic nucleus consists of protons and neutrons. At the nuclear scale and in the isospin formalism introduced by Heisenberg\,\cite{Hei32} and Wigner\,\cite{Wig37}, these are understood as two projections of the same particle, the nucleon. If isospin is conserved, the nucleon-nucleon interaction $V_{\rho \rho}$ would be expected to be charge symmetric ($V_{pp} = V_{nn}$) and charge independent ($V_{pn} = (V_{pp}+V_{nn})/2$)\,\cite{War06}. However, isospin symmetry will be broken by any component of the nuclear Hamiltonian which discriminates between protons and neutrons. One such component is already the Coulomb interaction, acting only between protons due to their electric charge, which will introduce binding-energy differences of several MeV in nuclei belonging to the same isospin multiplet. In addition to the Coulomb interaction, the nucleon-nucleon interaction itself breaks isospin symmetry. 

Recently, Ormand {\it et al.} quantified the contribution of the charge-symmetry breaking (CSB) part of the nucleon-nucleon interaction. They calculated the $c$ coefficients of the isobaric mass multiplet equation for triplets of isobaric nuclei in the $fp$ shell starting from three state-of-the-art, realistic nucleon-nucleon interactions\,\cite{Orm17a}. In these studies, they showed that each of the derived effective two-body CSB interactions gave different results for the $c$ coefficients and did not agree with the experimental data. On the one hand, the latter suggested a possibly smaller contribution of the CSB part of the interaction to the $c$ coefficients. Previously, similar conclusions were drawn by Gadea {\it et al.}\,\cite{Gad06a} who compared their experimental data on the $A = 54$ nuclei with shell-model calculations using a charge-dependent interaction based on the AV18 potential. On the other hand, (i) the deviations among the interactions suggested that either CSB of the nucleon-nucleon interaction is currently poorly known, that (ii) there is strong CSB in the three-nucleon interaction, or that (iii) there is a significant induced three-nucleon interaction arising from the renormalization procedure. It was, however, pointed out that the CSB contribution to the $c$ coefficient is almost independent of the order of renormalization suggesting that the CSB part of the interaction is predominantly of short range\,\cite{Orm17a}. The strong $J$ dependence of the CSB part found by Ormand {\it et al.}\,\cite{Orm17a} is in line with the work of \cite{Zuk02, Gad06a, Len18a} that studied triplet energy differences (TED) in the $fp$ shell. In those studies, an isotensor two-body matrix element $V_B^{(2),J}$ was empirically determined. It was found that the effect of this matrix element on the TED, needed to explain the experimental data, was as large as that of the Coulomb matrix element\,\cite{Len18a}. 
\\

An important benchmark system to understand charge-dependent contributions for nuclei between the doubly-magic $^{40}$Ca and $^{56}$Ni is the $T=1$ isospin triplet ($^{54}$Ni,$^{54}$Co,$^{54}$Fe). In contrast to the cross conjugate $A = 42$ nuclei\,\cite{Kut78a}, negligible cross-shell mixing with the $sd$ shell is expected\,\cite{Gad06a,Orm17a}. Therefore, rather pure $(1f_{7/2})^{-2}$ two-hole configurations should be observed for the yrast states allowing for a clear comparison to large-scale-shell-model (LSSM) calculations performed in the full $fp$-model space. Due to this dominant and simple configuration, large overlaps between the $(1f_{7/2})^{-1}$ $A=55$ ground and $(1f_{7/2})^{-2}$ $A=54$ yrast states are expected in one-nucleon knockout reactions. These reactions\,\cite{Han03,Gad08}, therefore, provide a selective way to populate the states of interest to study isospin-symmetry breaking in the $A=54$, $T=1$ triplet and to clearly identify the $(T = 1, J^{\pi} = 6^+)$ state of $^{54}$Co, which is presently still lacking clear evidence. A candidate for the $(T = 1, T_z = 0, J^{\pi} = 6^+)$ state has been previously reported in\,\cite{Rud10a}. On the tail of a strong $\gamma$-ray line of $^{53}$Fe with $E_{\gamma} = 2843$\,keV, Rudolph {\it et al.}\,\cite{Rud10a} observed an excess of counts, but without a clear peak shape, at $E_{\gamma} = 2782$\,keV in their normalized total projection of the $\gamma \gamma$ matrix obtained after an advanced particle gating procedure. They argued that this weak transition could correspond to the $6^+ \rightarrow 7^+_1$ transition and could, thus, establish the $(T = 1, T_z = 0, J^{\pi} = 6^+)$ state at 2979\,keV. The experimental TED of $+62$\,keV would, however, not follow the general trend of negative TED, which was observed for the $J^{\pi} = 6^+$ isospin-triplet states in this mass region\,\cite{Len18a}. 

\begin{figure}[t]
\centering
\includegraphics[width=1\linewidth]{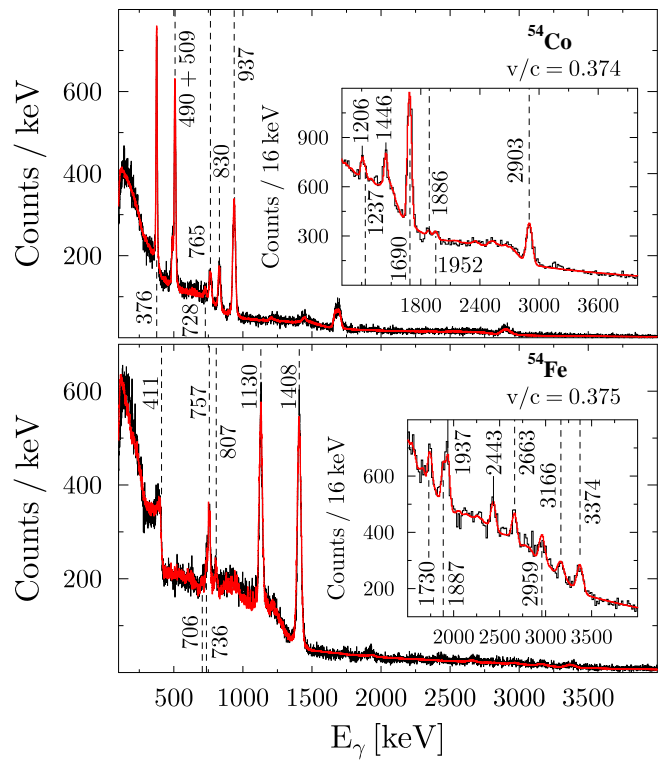}
\caption{\label{fig:spectra}{(color online) In-beam $\gamma$-ray singles
    spectra for $^{54}$Co (top) and $^{54}$Fe (bottom) in black compared to
    $\gamma$-ray spectra obtained from a GEANT4 simulation (red). Observed
    transitions are marked with dashed vertical lines and their corresponding
    transition energies. The background structures between $400-800$\,keV, seen on top of the smooth background, are caused by $\gamma$ rays emitted from stopped components and taken into account in the simulation. The structure at around 3.2\,MeV, seen in the inset for $^{54}$Co, did not unambiguously resemble a clear peak shape and was, therefore, omitted in the simulation. The placement of the transitions in the level schemes is summarized in Fig.\,\ref{fig:level}.}} 
 \end{figure}

This work reports on the identification of the sought-after $(T = 1, J^{\pi} = 6^+)$ state of $^{54}$Co via one-neutron knockout from $^{55}$Co but does not confirm the previous candidate\,\cite{Rud10a}. Isospin symmetry is discussed based on a comparison to the one-proton knockout populating $T=1$ states of $^{54}$Fe. The now complete experimental results on the $(1f_{7/2})^{-2}$ $A=54$ yrast states are compared to LSSM calculations in the full $fp$-model space. Furthermore, the full $J$ dependence of the CSB part, as predicted by Ormand\,{\it et al.}\,\cite{Orm17a}, is tested for the $T=1$ $(1f_{7/2})^{-2}$ yrast states using the theoretical results obtained with a realistic chiral effective field theory potential at next-to-next-to-next-to-leading order (N$^3$LO).

\begin{figure}[t]
\centering
\includegraphics[width=1\linewidth]{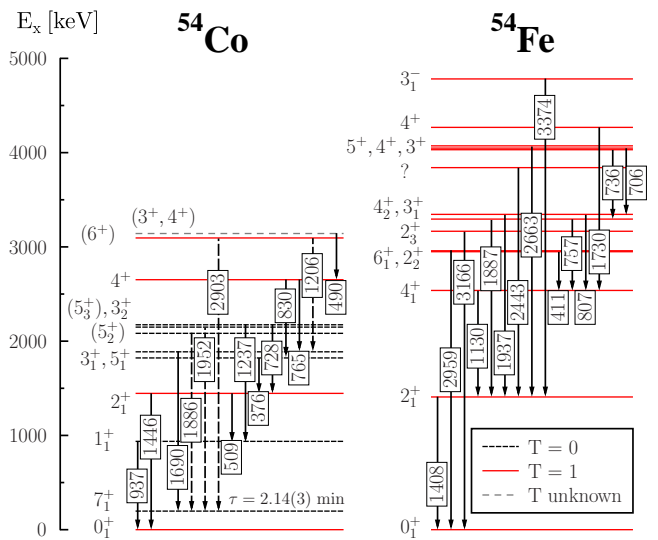}
\caption{\label{fig:level}{(color online) Observed level schemes of $^{54}$Co and $^{54}$Fe. Shown are $\gamma$-ray transitions which could be identified in the $\gamma$-ray spectra (compare Fig.\,\ref{fig:spectra}). $T = 0$ states are shown with shorther horizontally-dashed lines (black) and $T = 1$ with solid, horizontal lines (red). States with an uncertain isospin character are presented with longer horizontally-dashed lines (grey). Tentatively placed transitions are shown with vertically-dashed lines. For all states but the last two observed excited states in $^{54}$Co, spin-parity assignments were adopted from\,\cite{Dong14a}. For the latter, $J^{\pi}$ assignments are proposed based on a comparison to $^{54}$Fe.}} 
 \end{figure}

The one-nucleon knockout experiments were performed at the Coupled Cyclotron Facility of the National
Superconducting Cyclotron Laboratory (NSCL) at Michigan State
University\,\cite{NSCL}. The secondary $^{55}$Co beam was one component (27\,$\%$) of a secondary-beam cocktail produced from a 160\,MeV/u $^{58}$Ni primary beam in projectile
fragmentation on a thick 610 mg/cm$^2$ $^{9}$Be target. The A1900 fragment separator\,\cite{Mor03}, using a 300 mg/cm$^2$ Al degrader, was tuned to select the campaign's major fragment of interest, $^{56}$Ni, in flight. The secondary beam of interest for this work, $^{55}$Co, could be unambigiously distinguished from $^{56}$Ni (72\,$\%$) and the other small $^{54}$Fe contaminant (1\,$\%$) via the
time-of-flight difference measured between two plastic scintillators located at
the exit of the A1900 and the object position of the S800 analysis beam
line. Downstream, the secondary $^{9}$Be reaction target (188 mg/cm$^2$ thick) was located at the target position of the S800 spectrograph. 
The projectile-like reaction residues entering the S800 focal plane were identified event-by-event from their
energy loss and time of flight\,\cite{Baz03}. The inclusive cross sections $\sigma_{inc.}$ for the one-neutron and one-proton knockout from $^{55}$Co to bound final states of $^{54}$Co and $^{54}$Fe were deduced to be $39.0 \pm 0.4 \, (\mathrm{stat.}) \pm 2.8 \,(\mathrm{sys.})$\,mb and $141 \pm 3 \, (\mathrm{stat.}) \pm 16 \, (\mathrm{sys.})$\,mb, respectively. Systematic uncertainties
include the stability of the secondary beam composition, the choice of software gates, and corrections for acceptance losses in the tails of the residue
parallel momentum distributions due to the blocking of the unreacted beam in the focal plane. Only a change in magnetic rigidity of the S800 spectrograph was required to switch from the one-neutron to the one-proton knockout setting.

To detect the $\gamma$ rays emitted by the reaction residues in flight ($v/c \approx 0.4$), the reaction target was surrounded by
the SeGA array\,\cite{Mue01}. The 16 32-fold segmented High-Purity Germanium detectors were arranged in two rings with central
angles of 37$^{\circ}$ (7 detectors) and 90$^{\circ}$ (9 detectors) relative to
the beam axis. Event-by-event
Doppler reconstruction of the residues' $\gamma$-ray energies was performed based on the angle of the $\gamma$-ray
emission determined from the segment position
that registered the highest energy deposition. In Fig.\,\ref{fig:spectra}, the Doppler-corrected in-beam
$\gamma$-ray singles spectra in coincidence with the event-by-event identified
knockout residues $^{54}$Co and $^{54}$Fe are shown together with corresponding GEANT4 simulations\,\cite{ucsega}. The adopted lifetime $\tau = 1.76(3)$\,ns of the $6^+_1$ level in $^{54}$Fe was taken into account in this simulation (see 411\,keV $\gamma$ ray in the lower panel of Fig.\,\ref{fig:spectra}). This state is short-lived enough to be detected with SeGA. The $\gamma$-ray emission of this state will, however, predominantly not take place in the center of the array and, therefore, cause a low-energy tail and a centroid shifted with respect to the nominal transition energy and the Doppler correction being performed with respect to the mid-target emission. The $\gamma$ decay of the isomeric $7^+_1$ state of $^{54}$Co ($\tau = 2.14(3)$\,min)\,\cite{Dong14a} could not be detected with SeGA. This $\gamma$ decay will take place long after the residue has reached the focal plane of the S800 spectrograph.

\begin{figure}[t]
\centering
\includegraphics[width=1\linewidth]{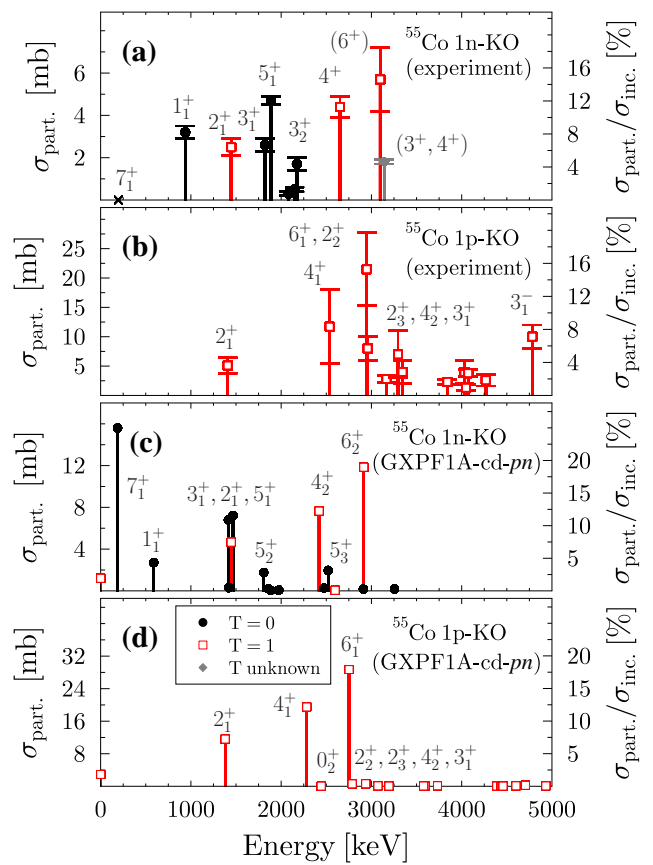}
\caption{\label{fig:partials}{(color online) Partial cross sections
    $\sigma_{\mathrm{part.}}$ determined for {\bf (a)} $^{54}$Co and {\bf (b)}
    $^{54}$Fe in comparison to {\bf (c), (d)} the theoretical cross
    sections. Only the first three states of each spin and predicted with $\sigma_{part.} \geq 0.03$\,mb are
    presented. $T = 0$ states are presented in black (full circles), $T=1$ states in red (open squares) and states with uncertain isospin character in grey (full diamonds). In addition, the partial
    cross sections relative to the inclusive cross section
    $\sigma_{\mathrm{inc.}}$ are shown, see second axis on the right. Only statistical
    uncertainties are given. No reduction factor $R_s$\,\cite{Gad08b,Tos14} has been applied for
    the comparison. The location of the $7^+_1$ in $^{54}$Co\,\cite{Dong14a} is indicated by a black cross in panel\,{\bf (a)}. As described in the text, this state was not observed because of its isomeric character. After subtraction, a partial cross section of 11(4)\,mb (27(11)$\%$) remains in $^{54}$Co. This would be in very good agreement with the theoretical expectations if it is mainly concentrated in the $7^+_1$ state and also accomodates the small partial cross section expected for the ground state. For further information on the theoretical cross sections, see the supplemental material\,\cite{suppl}}.} 
 \end{figure}

The level schemes of $^{54}$Co and $^{54}$Fe were largely known from previous work\,\cite{Dong14a}. In the odd-odd $N = Z = 27$ $^{54}$Co, excited states with isospin quantum numbers $T = 0$ and $T = 1$ are known to coexist at low excitation energy, see, {\it e.g.}, \cite{Sch00a, Bre02a, Rud10a} and references therein. Only the analog states with $T=1$ can be observed in $^{54}$Fe. The level schemes observed in one-nucleon knockout and the corresponding $\gamma$-ray transitions marked in Fig.\,\ref{fig:spectra} are shown in Fig.\,\ref{fig:level}. Spin-parity assignments, level and transition energies were adopted from evaluated data\,\cite{Dong14a} if not noted otherwise. No significant deviations from adopted transition energies were observed using our singles spectra of Fig.\,\ref{fig:spectra} and $\gamma\gamma$ coincidences utilized for more strongly populated states. The $\gamma$-ray yields were determined from the previously mentioned GEANT4 simulations\,\cite{ucsega}. From these $\gamma$-ray yields, the experimental partial cross sections $\sigma_{\mathrm{part.}}$, shown in Figs.\,\ref{fig:partials}\,{\bf (a)} and {\bf (b)} and feeding subtracted where possible, were calculated as described in, {\it e.g.},\,\cite{Gad08b, Str14, Gad16a} and references therein. 

Theoretical predictions for the partial cross sections, obtained from combining shell-model
spectroscopic factors with eikonal reaction theory\,\cite{Tos01} following the
approach outlined in detail in\,\cite{Gad08b,Tos14}, are shown in Figs.\,\ref{fig:partials}\,{\bf (c)} and {\bf (d)}. The LSSM calculations were performed in the full $fp$ shell using the GXPF1A-cd-$pn$ Hamiltonian with the effective isospin-conserving GXPF1A interaction from \cite{Hon02, Hon04, Hon05} and the charge-dependent (cd) Hamiltonian from\,\cite{Orm89}. The total wavefunctions were calculated in a proton-neutron basis
($pn$). Spectroscopic factors $C^2S(J^{\pi})$ were computed from the $^{55}$Co ground state, taking into account all possible contributions to the ground-state wavefunction, to bound final states with $J^{\pi}$ in $^{54}$Co ($^{55}$Co 1n-KO) and $^{54}$Fe ($^{55}$Co 1p-KO). These enter the knockout cross sections as described in more detail in the supplemental material\,\cite{suppl}. The final theoretical partial cross section is the sum of the individual knockout contributions from the $1f_{7/2}$, $2p_{3/2}$, $1f_{5/2}$, and $2p_{1/2}$ orbitals to a given final state, weighted according to their spectroscopic factors. The $1f_{7/2}$ knockout contribution is by far the largest. For the shell-model calculations, the computer code NuShellX was utilized\,\cite{NuX}. 

As mentioned in the introduction, the $(T=1, J^{\pi} = 6^+)$ state of $^{54}$Co has not yet been unambiguously identified. We will show that it can be clearly identified based on its $\gamma$-decay behavior and its strong population in one-neutron knockout from $^{55}$Co. For the other $T=1$ states, both experimental observables are well described within the LSSM calculations and, thus, provide unique fingerprints.  

\begin{figure}[t]
\centering
\includegraphics[width=.8\linewidth]{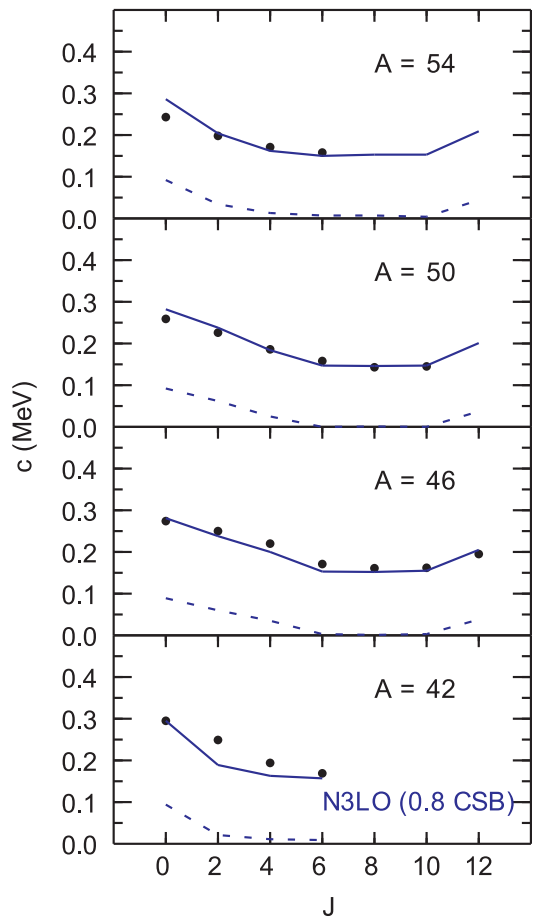}
\caption{\label{fig:c8}{(color online) First-order calculations for N$^3$LO with the CSB part multiplied by 0.8 compared to experiment as explained in Ref.\,\cite{Orm17a}. The black circles are the experimentally calculated $c$ coefficients using the latest atomic mass evaluation\,\cite{Wang17a}. This evaluation was not available when\,\cite{Orm17a} was published. The solid lines show the sum of the Coulomb and CSB contributions. The dashed lines show only the CSB contribution.}} 
 \end{figure}  

\begin{figure}[t]
\centering
\includegraphics[width=.8\linewidth]{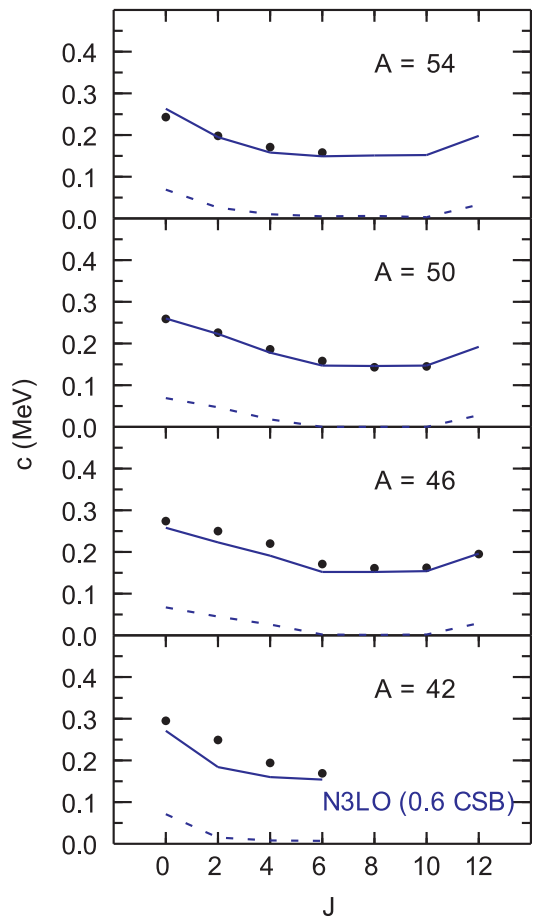}
\caption{\label{fig:c6}{(color online) Same as Fig.\,\ref{fig:c8} but with the CSB part multiplied by 0.6.}} 
 \end{figure}

The yrast states are expected to be strongly populated in one-nucleon knockout due to the large spectroscopic factors between the $^{55}$Co ground state and excited states in $^{54}$Co and $^{54}$Fe. The theoretical cross sections for the yrast states, that scale with the spectroscopic factors, increase with $J$ as seen in Figs.\,\ref{fig:partials}\,{\bf (c)},{\bf (d)}. Within uncertainties, this expectation and also the relative populations $\sigma_{\mathrm{part.}}/\sigma_{\mathrm{inc.}}$ agree well with the experimental results. The previously known $6^+_1$ state of $^{54}$Fe is indeed strongly populated, see Fig.\,\ref{fig:partials}\,{\bf (b)}, and identified via the $6^+_1 \rightarrow 4^+_1$ $\gamma$-ray transition. In contrast to $^{54}$Fe, the existence of the low-lying $T=0$ states in $^{54}$Co leads to fast $M1$ transitions from the $T=1$ states to the former. The main $\gamma$-decay branches of the $(T=1,J^{\pi}=2^+_1)$ and $(T=1,J^{\pi}=4^+)$ states lead to the $(T=0,J^{\pi}=1^+_1)$ (89(3)$\%$) for the former and $(T=0,J^{\pi}=3^+_1)$ (64(3)$\%$) as well as $(T=0,J^{\pi}=5^+_1)$ (36(2)$\%$) for the latter. The predicted $\gamma$-decay intensities of 98$\%$, 68$\%$ and 31$\%$ agree very well with experiment. For the $(T=1, J^{\pi}=6^+)$ state of $^{54}$Co, the present shell-model calculations predict $\gamma$-decay intensities of 80$\%$ and 17$\%$ to the $7^+_1$ and $5^+_1$ state, respectively, with $\gamma$-ray energies of 2727\,keV and 1448\,keV. Consequently and because of the expected strong population of the $(T=1, J^{\pi}=6^+)$ state in one-neutron knockout from $^{55}$Co (compare Fig.\,\ref{fig:partials}\,{\bf (c)}), a prominent $6^+ \rightarrow 7^+_1$ transition should be observed around 2.7\,MeV. The strongest and only $\gamma$-ray line in the relevant energy range appears at 2903(4)\,keV (compare Fig.\,\ref{fig:spectra} for $^{54}$Co). It has not been observed before. Assuming that this decay does indeed correspond to the $6^+ \rightarrow 7^+_1$ transition would establish a state at 3100(4)\,keV. If the predicted intensities are correct, a significant $6^+ \rightarrow 5^+_1$ transition at $E_{\gamma} = 1213(4)$\,keV would also be expected. A transition is observed at 1206(5)\,keV. This establishes an excited state in $^{54}$Co at 3097(6)\,keV with the expected decay pattern, i.e. 89(6)$\%$ to the $7^+_1$ and 11(2)$\%$ to the $5^+_1$ state. Furthermore, after correcting the corresponding $\gamma$-ray yields for $\gamma$-decay branching, a partial cross section is determined which fits well in the expected one-neutron knockout cross section pattern (compare Figs.\,\ref{fig:partials}\,{\bf (a)} and {\bf (c)}). We note, that there is no indication for a $\gamma$-ray transition with $E_{\gamma} = 2782$\,keV (compare Fig.\,\ref{fig:spectra}), which was previously proposed by Rudolph {\it et al.} to identify the $(T=1, J^{\pi} = 6^+)$ state in $^{54}$Co\,\cite{Rud10a}. The excited state at 3097(6)\,keV is, therefore, a strong and, in fact, the only possible candidate if isospin symmetry is not significantly broken in the $T=1$ triplet. This statement is supported by the very similar cross-section pattern observed for the $(T=1, J^{\pi} = 2^+_1)$ and $(T=1, J^{\pi} = 4^+)$ states in $^{54}$Co and $^{54}$Fe.  
\\ 

Having identified the so far strongest $(T=1,J^{\pi}=6^+)$ candidate in $^{54}$Co, we now turn to the discussion of the $c$ coefficients. These are shown in Figs.\,\ref{fig:c8} and \ref{fig:c6} for $A=42$, 46, 50, 54, and have been updated from\,\cite{Orm17a} with new results for $A = 54$ including the latest ground-state binding energies\,\cite{Wang17a} as well as the excitation energy for the new $(T=1, T_z = 0, J^{\pi} = 6^+)$ candidate. Equivalent information can be obtained by studying the TED\,\cite{Len18a, Ben07a}. As a reminder, the $c$ coefficients are $J$ dependent parameters of the isobaric mass multiplet equation (IMME)\,\cite{Wig57} and provide information on the isotensor component of the nuclear Hamiltonian. Information on the isovector component can be obtained by studying the $b$ coefficients or the corresponding mirror energy differences (MED) as done in, {\it e.g.}, Refs.\,\cite{Gad06a, Ben07a, Ben15}. As previously noted\,\cite{Orm17a}, adding 80$\%$ of the N$^3$LO CSB part to the first-order Coulomb contribution provided a fair description of the experimental $c$ coefficients in the $fp$ shell up to $^{56}$Ni. Larger deviations arise when getting closer to $^{40}$Ca due to cross-shell mixing with $sd$ shell-configurations for the low-lying positive-parity states (compare Fig.\,\ref{fig:c8}). Here, it is realized that lowering the CSB part even further, i.e. to 60$\%$ of the initially predicted value, provides an almost perfect agreement for $A=50$ and $A=54$ (compare Fig.\,\ref{fig:c6}). The $c$ coefficient for the new $(T=1, T_z = 0, J^{\pi} = 6^+)$ candidate agrees well with the predicted $J$ dependence. For completeness we note that the experimental ${\mathrm {TED}} = -174(6)$\,keV for $J=6$.
\\

In summary, we have performed the first one-nucleon knockout reactions from $^{55}$Co to further study isospin symmetry in the $A=54$, $T=1$ triplet. The new experimental data establishes the 3097(6)\,keV excited state of $^{54}$Co as the strongest candidate for the $(T=1, T_z = 0, J^{\pi} = 6^+)$ state. The state's energy was inferred from two transitions tentatively assigned to the $\gamma$ decays of the state to the $5^+_1$ and $7^+_1$ state, respectively. All observables are, however, in agreement with shell-model predictions and the experimentally measured partial cross section of the $(T=1, T_z = 1, J^{\pi} = 6^+)$ state in $^{54}$Fe. In general, the comparison of the experimental partial cross section pattern with theory confirms the $(1f_{7/2})^{-2}$ dominance of the wavefunctions and shows an overall good agreement. Nevertheless, even though missed feeding contributions and contributions from more complex excitation mechanisms cannot be excluded, yrare states are more strongly populated than predicted by the present LSSM calculations in both $^{54}$Co and $^{54}$Fe. The population of the $3^-_1$ state of $^{54}$Fe in one-proton knockout from $^{55}$Co might indicate that considering $(1f_{7/2})^{-1}(1s_{1/2})^1$ configurations at higher energies is important, as was also recently discussed in\,\cite{Spi19a}. However, as shown by the comparison of the $c$ coefficients, the influence of cross-shell mixing with the $sd$-shell configurations on the yrast states is negligible. An almost perfect agreement between experiment and theory is observed for $A=50 - 54$. The necessary reduction of the CSB part predicted with realistic nucleon-nucleon interactions remains a puzzle to be solved by nuclear theory.  
\\

This work was supported by the National Science Foundation under Grant No. PHY-1565546 (NSCL) and PHY-1811855, the DOE National Nuclear Security Administration through the Nuclear Science and Security Consortium under Award No. DE-NA0003180 as well as by the UK Science and Technology Facilities Council (STFC) through Research Grant No. ST/F005314/1, ST/P003885/1 and GR/T18486/01.


\bibliography{55Co}

\providecommand{\noopsort}[1]{}\providecommand{\singleletter}[1]{#1}%
\begin{thebibliography}{35}%
\makeatletter
\providecommand \@ifxundefined [1]{%
 \@ifx{#1\undefined}
}%
\providecommand \@ifnum [1]{%
 \ifnum #1\expandafter \@firstoftwo
 \else \expandafter \@secondoftwo
 \fi
}%
\providecommand \@ifx [1]{%
 \ifx #1\expandafter \@firstoftwo
 \else \expandafter \@secondoftwo
 \fi
}%
\providecommand \natexlab [1]{#1}%
\providecommand \enquote  [1]{``#1''}%
\providecommand \bibnamefont  [1]{#1}%
\providecommand \bibfnamefont [1]{#1}%
\providecommand \citenamefont [1]{#1}%
\providecommand \href@noop [0]{\@secondoftwo}%
\providecommand \href [0]{\begingroup \@sanitize@url \@href}%
\providecommand \@href[1]{\@@startlink{#1}\@@href}%
\providecommand \@@href[1]{\endgroup#1\@@endlink}%
\providecommand \@sanitize@url [0]{\catcode `\\12\catcode `\$12\catcode
  `\&12\catcode `\#12\catcode `\^12\catcode `\_12\catcode `\%12\relax}%
\providecommand \@@startlink[1]{}%
\providecommand \@@endlink[0]{}%
\providecommand \url  [0]{\begingroup\@sanitize@url \@url }%
\providecommand \@url [1]{\endgroup\@href {#1}{\urlprefix }}%
\providecommand \urlprefix  [0]{URL }%
\providecommand \Eprint [0]{\href }%
\providecommand \doibase [0]{http://dx.doi.org/}%
\providecommand \selectlanguage [0]{\@gobble}%
\providecommand \bibinfo  [0]{\@secondoftwo}%
\providecommand \bibfield  [0]{\@secondoftwo}%
\providecommand \translation [1]{[#1]}%
\providecommand \BibitemOpen [0]{}%
\providecommand \bibitemStop [0]{}%
\providecommand \bibitemNoStop [0]{.\EOS\space}%
\providecommand \EOS [0]{\spacefactor3000\relax}%
\providecommand \BibitemShut  [1]{\csname bibitem#1\endcsname}%
\let\auto@bib@innerbib\@empty
\bibitem [{\citenamefont {Heisenberg}(1932)}]{Hei32}%
  \BibitemOpen
  \bibfield  {author} {\bibinfo {author} {\bibfnamefont {W.}~\bibnamefont
  {Heisenberg}},\ }\href {\doibase 10.1007/BF01342433} {\bibfield  {journal}
  {\bibinfo  {journal} {Zeitschrift f{\"u}r Physik}\ }\textbf {\bibinfo
  {volume} {77}},\ \bibinfo {pages} {1} (\bibinfo {year} {1932})}\BibitemShut
  {NoStop}%
\bibitem [{\citenamefont {Wigner}(1937)}]{Wig37}%
  \BibitemOpen
  \bibfield  {author} {\bibinfo {author} {\bibfnamefont {E.}~\bibnamefont
  {Wigner}},\ }\href {\doibase 10.1103/PhysRev.51.106} {\bibfield  {journal}
  {\bibinfo  {journal} {Phys. Rev.}\ }\textbf {\bibinfo {volume} {51}},\
  \bibinfo {pages} {106} (\bibinfo {year} {1937})}\BibitemShut {NoStop}%
\bibitem [{\citenamefont {Warner}\ \emph {et~al.}(2006)\citenamefont {Warner},
  \citenamefont {Bentley},\ and\ \citenamefont {van Isacker}}]{War06}%
  \BibitemOpen
  \bibfield  {author} {\bibinfo {author} {\bibfnamefont {D.~D.}\ \bibnamefont
  {Warner}}, \bibinfo {author} {\bibfnamefont {M.~A.}\ \bibnamefont {Bentley}},
  \ and\ \bibinfo {author} {\bibfnamefont {P.}~\bibnamefont {van Isacker}},\
  }\href@noop {} {\bibfield  {journal} {\bibinfo  {journal} {Nature Physics}\
  }\textbf {\bibinfo {volume} {2}},\ \bibinfo {pages} {311} (\bibinfo {year}
  {2006})}\BibitemShut {NoStop}%
\bibitem [{\citenamefont {Ormand}\ \emph {et~al.}(2017)\citenamefont {Ormand},
  \citenamefont {Brown},\ and\ \citenamefont {Hjorth-Jensen}}]{Orm17a}%
  \BibitemOpen
  \bibfield  {author} {\bibinfo {author} {\bibfnamefont {W.~E.}\ \bibnamefont
  {Ormand}}, \bibinfo {author} {\bibfnamefont {B.~A.}\ \bibnamefont {Brown}}, \
  and\ \bibinfo {author} {\bibfnamefont {M.}~\bibnamefont {Hjorth-Jensen}},\
  }\href {\doibase 10.1103/PhysRevC.96.024323} {\bibfield  {journal} {\bibinfo
  {journal} {Phys. Rev. C}\ }\textbf {\bibinfo {volume} {96}},\ \bibinfo
  {pages} {024323} (\bibinfo {year} {2017})}\BibitemShut {NoStop}%
\bibitem [{\citenamefont {Gadea}\ \emph {et~al.}(2006)\citenamefont {Gadea},
  \citenamefont {Lenzi}, \citenamefont {Lunardi}, \citenamefont
  {M\ifmmode~\u{a}\else \u{a}\fi{}rginean}, \citenamefont {Zuker},
  \citenamefont {de~Angelis}, \citenamefont {Axiotis}, \citenamefont
  {Mart\'{\i}nez}, \citenamefont {Napoli}, \citenamefont {Farnea},
  \citenamefont {Menegazzo}, \citenamefont {Pavan}, \citenamefont {Ur},
  \citenamefont {Bazzacco}, \citenamefont {Venturelli}, \citenamefont
  {Kleinheinz}, \citenamefont {Bednarczyk}, \citenamefont {Curien},
  \citenamefont {Dorvaux}, \citenamefont {Nyberg}, \citenamefont {Grawe},
  \citenamefont {G\'orska}, \citenamefont {Palacz}, \citenamefont {Lagergren},
  \citenamefont {Milechina}, \citenamefont {Ekman}, \citenamefont {Rudolph},
  \citenamefont {Andreoiu}, \citenamefont {Bentley}, \citenamefont {Gelletly},
  \citenamefont {Rubio}, \citenamefont {Algora}, \citenamefont {Nacher},
  \citenamefont {Caballero}, \citenamefont {Trotta},\ and\ \citenamefont
  {Moszy\ifmmode~\acute{n}\else \'{n}\fi{}ski}}]{Gad06a}%
  \BibitemOpen
  \bibfield  {author} {\bibinfo {author} {\bibfnamefont {A.}~\bibnamefont
  {Gadea}}, \bibinfo {author} {\bibfnamefont {S.~M.}\ \bibnamefont {Lenzi}},
  \bibinfo {author} {\bibfnamefont {S.}~\bibnamefont {Lunardi}}, \bibinfo
  {author} {\bibfnamefont {N.}~\bibnamefont {M\ifmmode~\u{a}\else
  \u{a}\fi{}rginean}}, \bibinfo {author} {\bibfnamefont {A.~P.}\ \bibnamefont
  {Zuker}}, \bibinfo {author} {\bibfnamefont {G.}~\bibnamefont {de~Angelis}},
  \bibinfo {author} {\bibfnamefont {M.}~\bibnamefont {Axiotis}}, \bibinfo
  {author} {\bibfnamefont {T.}~\bibnamefont {Mart\'{\i}nez}}, \bibinfo {author}
  {\bibfnamefont {D.~R.}\ \bibnamefont {Napoli}}, \bibinfo {author}
  {\bibfnamefont {E.}~\bibnamefont {Farnea}}, \bibinfo {author} {\bibfnamefont
  {R.}~\bibnamefont {Menegazzo}}, \bibinfo {author} {\bibfnamefont
  {P.}~\bibnamefont {Pavan}}, \bibinfo {author} {\bibfnamefont {C.~A.}\
  \bibnamefont {Ur}}, \bibinfo {author} {\bibfnamefont {D.}~\bibnamefont
  {Bazzacco}}, \bibinfo {author} {\bibfnamefont {R.}~\bibnamefont
  {Venturelli}}, \bibinfo {author} {\bibfnamefont {P.}~\bibnamefont
  {Kleinheinz}}, \bibinfo {author} {\bibfnamefont {P.}~\bibnamefont
  {Bednarczyk}}, \bibinfo {author} {\bibfnamefont {D.}~\bibnamefont {Curien}},
  \bibinfo {author} {\bibfnamefont {O.}~\bibnamefont {Dorvaux}}, \bibinfo
  {author} {\bibfnamefont {J.}~\bibnamefont {Nyberg}}, \bibinfo {author}
  {\bibfnamefont {H.}~\bibnamefont {Grawe}}, \bibinfo {author} {\bibfnamefont
  {M.}~\bibnamefont {G\'orska}}, \bibinfo {author} {\bibfnamefont
  {M.}~\bibnamefont {Palacz}}, \bibinfo {author} {\bibfnamefont
  {K.}~\bibnamefont {Lagergren}}, \bibinfo {author} {\bibfnamefont
  {L.}~\bibnamefont {Milechina}}, \bibinfo {author} {\bibfnamefont
  {J.}~\bibnamefont {Ekman}}, \bibinfo {author} {\bibfnamefont
  {D.}~\bibnamefont {Rudolph}}, \bibinfo {author} {\bibfnamefont
  {C.}~\bibnamefont {Andreoiu}}, \bibinfo {author} {\bibfnamefont {M.~A.}\
  \bibnamefont {Bentley}}, \bibinfo {author} {\bibfnamefont {W.}~\bibnamefont
  {Gelletly}}, \bibinfo {author} {\bibfnamefont {B.}~\bibnamefont {Rubio}},
  \bibinfo {author} {\bibfnamefont {A.}~\bibnamefont {Algora}}, \bibinfo
  {author} {\bibfnamefont {E.}~\bibnamefont {Nacher}}, \bibinfo {author}
  {\bibfnamefont {L.}~\bibnamefont {Caballero}}, \bibinfo {author}
  {\bibfnamefont {M.}~\bibnamefont {Trotta}}, \ and\ \bibinfo {author}
  {\bibfnamefont {M.}~\bibnamefont {Moszy\ifmmode~\acute{n}\else
  \'{n}\fi{}ski}},\ }\href {\doibase 10.1103/PhysRevLett.97.152501} {\bibfield
  {journal} {\bibinfo  {journal} {Phys. Rev. Lett.}\ }\textbf {\bibinfo
  {volume} {97}},\ \bibinfo {pages} {152501} (\bibinfo {year}
  {2006})}\BibitemShut {NoStop}%
\bibitem [{\citenamefont {Zuker}\ \emph {et~al.}(2002)\citenamefont {Zuker},
  \citenamefont {Lenzi}, \citenamefont {Mart\'{\i}nez-Pinedo},\ and\
  \citenamefont {Poves}}]{Zuk02}%
  \BibitemOpen
  \bibfield  {author} {\bibinfo {author} {\bibfnamefont {A.~P.}\ \bibnamefont
  {Zuker}}, \bibinfo {author} {\bibfnamefont {S.~M.}\ \bibnamefont {Lenzi}},
  \bibinfo {author} {\bibfnamefont {G.}~\bibnamefont {Mart\'{\i}nez-Pinedo}}, \
  and\ \bibinfo {author} {\bibfnamefont {A.}~\bibnamefont {Poves}},\ }\href
  {\doibase 10.1103/PhysRevLett.89.142502} {\bibfield  {journal} {\bibinfo
  {journal} {Phys. Rev. Lett.}\ }\textbf {\bibinfo {volume} {89}},\ \bibinfo
  {pages} {142502} (\bibinfo {year} {2002})}\BibitemShut {NoStop}%
\bibitem [{\citenamefont {Lenzi}\ \emph {et~al.}(2018)\citenamefont {Lenzi},
  \citenamefont {Bentley}, \citenamefont {Lau},\ and\ \citenamefont
  {Diget}}]{Len18a}%
  \BibitemOpen
  \bibfield  {author} {\bibinfo {author} {\bibfnamefont {S.~M.}\ \bibnamefont
  {Lenzi}}, \bibinfo {author} {\bibfnamefont {M.~A.}\ \bibnamefont {Bentley}},
  \bibinfo {author} {\bibfnamefont {R.}~\bibnamefont {Lau}}, \ and\ \bibinfo
  {author} {\bibfnamefont {C.~A.}\ \bibnamefont {Diget}},\ }\href {\doibase
  10.1103/PhysRevC.98.054322} {\bibfield  {journal} {\bibinfo  {journal} {Phys.
  Rev. C}\ }\textbf {\bibinfo {volume} {98}},\ \bibinfo {pages} {054322}
  (\bibinfo {year} {2018})}\BibitemShut {NoStop}%
\bibitem [{\citenamefont {Kutschera}\ \emph {et~al.}(1978)\citenamefont
  {Kutschera}, \citenamefont {Brown},\ and\ \citenamefont {Ogawa}}]{Kut78a}%
  \BibitemOpen
  \bibfield  {author} {\bibinfo {author} {\bibfnamefont {W.}~\bibnamefont
  {Kutschera}}, \bibinfo {author} {\bibfnamefont {B.~A.}\ \bibnamefont
  {Brown}}, \ and\ \bibinfo {author} {\bibfnamefont {K.}~\bibnamefont
  {Ogawa}},\ }\href {\doibase 10.1007/BF02724440} {\bibfield  {journal}
  {\bibinfo  {journal} {La Rivista del Nuovo Cimento (1978-1999)}\ }\textbf
  {\bibinfo {volume} {1}},\ \bibinfo {pages} {1} (\bibinfo {year}
  {1978})}\BibitemShut {NoStop}%
\bibitem [{\citenamefont {Hansen}\ and\ \citenamefont
  {Tostevin}(2003)}]{Han03}%
  \BibitemOpen
  \bibfield  {author} {\bibinfo {author} {\bibfnamefont {P.}~\bibnamefont
  {Hansen}}\ and\ \bibinfo {author} {\bibfnamefont {J.~A.}\ \bibnamefont
  {Tostevin}},\ }\href@noop {} {\bibfield  {journal} {\bibinfo  {journal}
  {Annu. Rev. Nucl. Part. Sci.}\ }\textbf {\bibinfo {volume} {53}},\ \bibinfo
  {pages} {219} (\bibinfo {year} {2003})}\BibitemShut {NoStop}%
\bibitem [{\citenamefont {Gade}\ and\ \citenamefont
  {Glasmacher}(2008)}]{Gad08}%
  \BibitemOpen
  \bibfield  {author} {\bibinfo {author} {\bibfnamefont {A.}~\bibnamefont
  {Gade}}\ and\ \bibinfo {author} {\bibfnamefont {T.}~\bibnamefont
  {Glasmacher}},\ }\href@noop {} {\bibfield  {journal} {\bibinfo  {journal}
  {Prog. Part. Nucl. Phys.}\ }\textbf {\bibinfo {volume} {60}},\ \bibinfo
  {pages} {161} (\bibinfo {year} {2008})}\BibitemShut {NoStop}%
\bibitem [{\citenamefont {Rudolph}\ \emph {et~al.}(2010)\citenamefont
  {Rudolph}, \citenamefont {Andersson}, \citenamefont {Bengtsson},
  \citenamefont {Ekman}, \citenamefont {Erten}, \citenamefont {Fahlander},
  \citenamefont {Johansson}, \citenamefont {Ragnarsson}, \citenamefont
  {Andreoiu}, \citenamefont {Bentley}, \citenamefont {Carpenter}, \citenamefont
  {Charity}, \citenamefont {Clark}, \citenamefont {Fallon}, \citenamefont
  {Macchiavelli}, \citenamefont {Reviol}, \citenamefont {Sarantites},
  \citenamefont {Seweryniak}, \citenamefont {Svensson},\ and\ \citenamefont
  {Williams}}]{Rud10a}%
  \BibitemOpen
  \bibfield  {author} {\bibinfo {author} {\bibfnamefont {D.}~\bibnamefont
  {Rudolph}}, \bibinfo {author} {\bibfnamefont {L.-L.}\ \bibnamefont
  {Andersson}}, \bibinfo {author} {\bibfnamefont {R.}~\bibnamefont
  {Bengtsson}}, \bibinfo {author} {\bibfnamefont {J.}~\bibnamefont {Ekman}},
  \bibinfo {author} {\bibfnamefont {O.}~\bibnamefont {Erten}}, \bibinfo
  {author} {\bibfnamefont {C.}~\bibnamefont {Fahlander}}, \bibinfo {author}
  {\bibfnamefont {E.~K.}\ \bibnamefont {Johansson}}, \bibinfo {author}
  {\bibfnamefont {I.}~\bibnamefont {Ragnarsson}}, \bibinfo {author}
  {\bibfnamefont {C.}~\bibnamefont {Andreoiu}}, \bibinfo {author}
  {\bibfnamefont {M.~A.}\ \bibnamefont {Bentley}}, \bibinfo {author}
  {\bibfnamefont {M.~P.}\ \bibnamefont {Carpenter}}, \bibinfo {author}
  {\bibfnamefont {R.~J.}\ \bibnamefont {Charity}}, \bibinfo {author}
  {\bibfnamefont {R.~M.}\ \bibnamefont {Clark}}, \bibinfo {author}
  {\bibfnamefont {P.}~\bibnamefont {Fallon}}, \bibinfo {author} {\bibfnamefont
  {A.~O.}\ \bibnamefont {Macchiavelli}}, \bibinfo {author} {\bibfnamefont
  {W.}~\bibnamefont {Reviol}}, \bibinfo {author} {\bibfnamefont {D.~G.}\
  \bibnamefont {Sarantites}}, \bibinfo {author} {\bibfnamefont
  {D.}~\bibnamefont {Seweryniak}}, \bibinfo {author} {\bibfnamefont {C.~E.}\
  \bibnamefont {Svensson}}, \ and\ \bibinfo {author} {\bibfnamefont {S.~J.}\
  \bibnamefont {Williams}},\ }\href {\doibase 10.1103/PhysRevC.82.054309}
  {\bibfield  {journal} {\bibinfo  {journal} {Phys. Rev. C}\ }\textbf {\bibinfo
  {volume} {82}},\ \bibinfo {pages} {054309} (\bibinfo {year}
  {2010})}\BibitemShut {NoStop}%
\bibitem [{\citenamefont {Dong}\ and\ \citenamefont {Junde}(2014)}]{Dong14a}%
  \BibitemOpen
  \bibfield  {author} {\bibinfo {author} {\bibfnamefont {Y.}~\bibnamefont
  {Dong}}\ and\ \bibinfo {author} {\bibfnamefont {H.}~\bibnamefont {Junde}},\
  }\href {\doibase https://doi.org/10.1016/j.nds.2014.09.001} {\bibfield
  {journal} {\bibinfo  {journal} {Nuclear Data Sheets}\ }\textbf {\bibinfo
  {volume} {121}},\ \bibinfo {pages} {1 } (\bibinfo {year} {2014})}\BibitemShut
  {NoStop}%
\bibitem [{\citenamefont {Gade}\ and\ \citenamefont {Sherrill}(2016)}]{NSCL}%
  \BibitemOpen
  \bibfield  {author} {\bibinfo {author} {\bibfnamefont {A.}~\bibnamefont
  {Gade}}\ and\ \bibinfo {author} {\bibfnamefont {B.}~\bibnamefont
  {Sherrill}},\ }\href@noop {} {\bibfield  {journal} {\bibinfo  {journal}
  {Physica Scripta}\ }\textbf {\bibinfo {volume} {91}},\ \bibinfo {pages}
  {053003} (\bibinfo {year} {2016})}\BibitemShut {NoStop}%
\bibitem [{\citenamefont {Morrissey}\ \emph {et~al.}(2003)\citenamefont
  {Morrissey}, \citenamefont {Sherrill}, \citenamefont {Steiner}, \citenamefont
  {Stolz},\ and\ \citenamefont {Wiedenhoever}}]{Mor03}%
  \BibitemOpen
  \bibfield  {author} {\bibinfo {author} {\bibfnamefont {D.}~\bibnamefont
  {Morrissey}}, \bibinfo {author} {\bibfnamefont {B.}~\bibnamefont {Sherrill}},
  \bibinfo {author} {\bibfnamefont {M.}~\bibnamefont {Steiner}}, \bibinfo
  {author} {\bibfnamefont {A.}~\bibnamefont {Stolz}}, \ and\ \bibinfo {author}
  {\bibfnamefont {I.}~\bibnamefont {Wiedenhoever}},\ }\href@noop {} {\bibfield
  {journal} {\bibinfo  {journal} {Nucl. Instr. and Meth. B}\ }\textbf {\bibinfo
  {volume} {204}},\ \bibinfo {pages} {90} (\bibinfo {year} {2003})}\BibitemShut
  {NoStop}%
\bibitem [{\citenamefont {Bazin}\ \emph {et~al.}(2003)\citenamefont {Bazin},
  \citenamefont {Caggiano}, \citenamefont {Sherrill}, \citenamefont {Yurkon},\
  and\ \citenamefont {Zeller}}]{Baz03}%
  \BibitemOpen
  \bibfield  {author} {\bibinfo {author} {\bibfnamefont {D.}~\bibnamefont
  {Bazin}}, \bibinfo {author} {\bibfnamefont {J.}~\bibnamefont {Caggiano}},
  \bibinfo {author} {\bibfnamefont {B.}~\bibnamefont {Sherrill}}, \bibinfo
  {author} {\bibfnamefont {J.}~\bibnamefont {Yurkon}}, \ and\ \bibinfo {author}
  {\bibfnamefont {A.}~\bibnamefont {Zeller}},\ }\href@noop {} {\bibfield
  {journal} {\bibinfo  {journal} {Nucl. Instr. and Meth. B}\ }\textbf {\bibinfo
  {volume} {204}},\ \bibinfo {pages} {629} (\bibinfo {year}
  {2003})}\BibitemShut {NoStop}%
\bibitem [{\citenamefont {Mueller}\ \emph {et~al.}(2001)\citenamefont
  {Mueller}, \citenamefont {Church}, \citenamefont {Glasmacher}, \citenamefont
  {Gutknecht}, \citenamefont {Hackman}, \citenamefont {Hansen}, \citenamefont
  {Hu}, \citenamefont {Miller},\ and\ \citenamefont {Quirin}}]{Mue01}%
  \BibitemOpen
  \bibfield  {author} {\bibinfo {author} {\bibfnamefont {W.}~\bibnamefont
  {Mueller}}, \bibinfo {author} {\bibfnamefont {J.}~\bibnamefont {Church}},
  \bibinfo {author} {\bibfnamefont {T.}~\bibnamefont {Glasmacher}}, \bibinfo
  {author} {\bibfnamefont {D.}~\bibnamefont {Gutknecht}}, \bibinfo {author}
  {\bibfnamefont {G.}~\bibnamefont {Hackman}}, \bibinfo {author} {\bibfnamefont
  {P.}~\bibnamefont {Hansen}}, \bibinfo {author} {\bibfnamefont
  {Z.}~\bibnamefont {Hu}}, \bibinfo {author} {\bibfnamefont {K.}~\bibnamefont
  {Miller}}, \ and\ \bibinfo {author} {\bibfnamefont {P.}~\bibnamefont
  {Quirin}},\ }\href@noop {} {\bibfield  {journal} {\bibinfo  {journal} {Nucl.
  Instr. and Meth. A}\ }\textbf {\bibinfo {volume} {466}},\ \bibinfo {pages}
  {492} (\bibinfo {year} {2001})}\BibitemShut {NoStop}%
\bibitem [{ucs()}]{ucsega}%
  \BibitemOpen
  \href@noop {} {}\bibinfo {howpublished} {UCSeGA GEANT4, L. A. Riley, Ursinus
  College, unpublished.}\BibitemShut {Stop}%
\bibitem [{\citenamefont {Gade}\ \emph {et~al.}(2008)\citenamefont {Gade},
  \citenamefont {Adrich}, \citenamefont {Bazin}, \citenamefont {Bowen},
  \citenamefont {Brown}, \citenamefont {Campbell}, \citenamefont {Cook},
  \citenamefont {Glasmacher}, \citenamefont {Hansen}, \citenamefont {Hosier},
  \citenamefont {McDaniel}, \citenamefont {McGlinchery}, \citenamefont
  {Obertelli}, \citenamefont {Siwek}, \citenamefont {Riley}, \citenamefont
  {Tostevin},\ and\ \citenamefont {Weisshaar}}]{Gad08b}%
  \BibitemOpen
  \bibfield  {author} {\bibinfo {author} {\bibfnamefont {A.}~\bibnamefont
  {Gade}}, \bibinfo {author} {\bibfnamefont {P.}~\bibnamefont {Adrich}},
  \bibinfo {author} {\bibfnamefont {D.}~\bibnamefont {Bazin}}, \bibinfo
  {author} {\bibfnamefont {M.~D.}\ \bibnamefont {Bowen}}, \bibinfo {author}
  {\bibfnamefont {B.~A.}\ \bibnamefont {Brown}}, \bibinfo {author}
  {\bibfnamefont {C.~M.}\ \bibnamefont {Campbell}}, \bibinfo {author}
  {\bibfnamefont {J.~M.}\ \bibnamefont {Cook}}, \bibinfo {author}
  {\bibfnamefont {T.}~\bibnamefont {Glasmacher}}, \bibinfo {author}
  {\bibfnamefont {P.~G.}\ \bibnamefont {Hansen}}, \bibinfo {author}
  {\bibfnamefont {K.}~\bibnamefont {Hosier}}, \bibinfo {author} {\bibfnamefont
  {S.}~\bibnamefont {McDaniel}}, \bibinfo {author} {\bibfnamefont
  {D.}~\bibnamefont {McGlinchery}}, \bibinfo {author} {\bibfnamefont
  {A.}~\bibnamefont {Obertelli}}, \bibinfo {author} {\bibfnamefont
  {K.}~\bibnamefont {Siwek}}, \bibinfo {author} {\bibfnamefont {L.~A.}\
  \bibnamefont {Riley}}, \bibinfo {author} {\bibfnamefont {J.~A.}\ \bibnamefont
  {Tostevin}}, \ and\ \bibinfo {author} {\bibfnamefont {D.}~\bibnamefont
  {Weisshaar}},\ }\href@noop {} {\bibfield  {journal} {\bibinfo  {journal}
  {Phys. Rev. C}\ }\textbf {\bibinfo {volume} {77}},\ \bibinfo {pages} {044306}
  (\bibinfo {year} {2008})}\BibitemShut {NoStop}%
\bibitem [{\citenamefont {Tostevin}\ and\ \citenamefont {Gade}(2014)}]{Tos14}%
  \BibitemOpen
  \bibfield  {author} {\bibinfo {author} {\bibfnamefont {J.~A.}\ \bibnamefont
  {Tostevin}}\ and\ \bibinfo {author} {\bibfnamefont {A.}~\bibnamefont
  {Gade}},\ }\href@noop {} {\bibfield  {journal} {\bibinfo  {journal} {Phys.
  Rev. C}\ }\textbf {\bibinfo {volume} {90}},\ \bibinfo {pages} {057602}
  (\bibinfo {year} {2014})}\BibitemShut {NoStop}%
\bibitem [{sup()}]{suppl}%
  \BibitemOpen
  \href@noop {} {}\bibinfo {howpublished} {See Supplemental Material at
  http://link.aps.org/supplemental/10.1103/yyy.zz.xxxxxx for full details of
  the reaction model parameters, shell-model spectroscopy, and calculated cross
  sections.}\BibitemShut {Stop}%
\bibitem [{\citenamefont {Schneider}\ \emph {et~al.}(2000)\citenamefont
  {Schneider}, \citenamefont {Lisetskiy}, \citenamefont {Frie\ss{}ner},
  \citenamefont {Jolos}, \citenamefont {Pietralla}, \citenamefont {Schmidt},
  \citenamefont {Weisshaar},\ and\ \citenamefont {von Brentano}}]{Sch00a}%
  \BibitemOpen
  \bibfield  {author} {\bibinfo {author} {\bibfnamefont {I.}~\bibnamefont
  {Schneider}}, \bibinfo {author} {\bibfnamefont {A.~F.}\ \bibnamefont
  {Lisetskiy}}, \bibinfo {author} {\bibfnamefont {C.}~\bibnamefont
  {Frie\ss{}ner}}, \bibinfo {author} {\bibfnamefont {R.~V.}\ \bibnamefont
  {Jolos}}, \bibinfo {author} {\bibfnamefont {N.}~\bibnamefont {Pietralla}},
  \bibinfo {author} {\bibfnamefont {A.}~\bibnamefont {Schmidt}}, \bibinfo
  {author} {\bibfnamefont {D.}~\bibnamefont {Weisshaar}}, \ and\ \bibinfo
  {author} {\bibfnamefont {P.}~\bibnamefont {von Brentano}},\ }\href {\doibase
  10.1103/PhysRevC.61.044312} {\bibfield  {journal} {\bibinfo  {journal} {Phys.
  Rev. C}\ }\textbf {\bibinfo {volume} {61}},\ \bibinfo {pages} {044312}
  (\bibinfo {year} {2000})}\BibitemShut {NoStop}%
\bibitem [{\citenamefont {von Brentano}\ \emph {et~al.}(2002)\citenamefont {von
  Brentano}, \citenamefont {Frießner}, \citenamefont {Jolos}, \citenamefont
  {Lisetskiy}, \citenamefont {Schmidt}, \citenamefont {Schneider},
  \citenamefont {Pietralla}, \citenamefont {Sebe},\ and\ \citenamefont
  {Otsuka}}]{Bre02a}%
  \BibitemOpen
  \bibfield  {author} {\bibinfo {author} {\bibfnamefont {P.}~\bibnamefont {von
  Brentano}}, \bibinfo {author} {\bibfnamefont {C.}~\bibnamefont {Frießner}},
  \bibinfo {author} {\bibfnamefont {R.~V.}\ \bibnamefont {Jolos}}, \bibinfo
  {author} {\bibfnamefont {A.~F.}\ \bibnamefont {Lisetskiy}}, \bibinfo {author}
  {\bibfnamefont {A.}~\bibnamefont {Schmidt}}, \bibinfo {author} {\bibfnamefont
  {I.}~\bibnamefont {Schneider}}, \bibinfo {author} {\bibfnamefont
  {N.}~\bibnamefont {Pietralla}}, \bibinfo {author} {\bibfnamefont
  {T.}~\bibnamefont {Sebe}}, \ and\ \bibinfo {author} {\bibfnamefont
  {T.}~\bibnamefont {Otsuka}},\ }\href {\doibase
  https://doi.org/10.1016/S0375-9474(02)00772-8} {\bibfield  {journal}
  {\bibinfo  {journal} {Nucl. Phys. A}\ }\textbf {\bibinfo {volume} {704}},\
  \bibinfo {pages} {115} (\bibinfo {year} {2002})}\BibitemShut {NoStop}%
\bibitem [{\citenamefont {Stroberg}\ \emph {et~al.}(2014)\citenamefont
  {Stroberg}, \citenamefont {Gade}, \citenamefont {Tostevin}, \citenamefont
  {Bader}, \citenamefont {Baugher}, \citenamefont {Bazin}, \citenamefont
  {Berryman}, \citenamefont {Brown}, \citenamefont {Campbell}, \citenamefont
  {Kemper}, \citenamefont {Langer}, \citenamefont {Lunderberg}, \citenamefont
  {Lemasson}, \citenamefont {Noji}, \citenamefont {Recchia}, \citenamefont
  {Walz}, \citenamefont {Weisshaar},\ and\ \citenamefont {Williams}}]{Str14}%
  \BibitemOpen
  \bibfield  {author} {\bibinfo {author} {\bibfnamefont {S.~R.}\ \bibnamefont
  {Stroberg}}, \bibinfo {author} {\bibfnamefont {A.}~\bibnamefont {Gade}},
  \bibinfo {author} {\bibfnamefont {J.~A.}\ \bibnamefont {Tostevin}}, \bibinfo
  {author} {\bibfnamefont {V.~M.}\ \bibnamefont {Bader}}, \bibinfo {author}
  {\bibfnamefont {T.}~\bibnamefont {Baugher}}, \bibinfo {author} {\bibfnamefont
  {D.}~\bibnamefont {Bazin}}, \bibinfo {author} {\bibfnamefont {J.~S.}\
  \bibnamefont {Berryman}}, \bibinfo {author} {\bibfnamefont {B.~A.}\
  \bibnamefont {Brown}}, \bibinfo {author} {\bibfnamefont {C.~M.}\ \bibnamefont
  {Campbell}}, \bibinfo {author} {\bibfnamefont {K.~W.}\ \bibnamefont
  {Kemper}}, \bibinfo {author} {\bibfnamefont {C.}~\bibnamefont {Langer}},
  \bibinfo {author} {\bibfnamefont {E.}~\bibnamefont {Lunderberg}}, \bibinfo
  {author} {\bibfnamefont {A.}~\bibnamefont {Lemasson}}, \bibinfo {author}
  {\bibfnamefont {S.}~\bibnamefont {Noji}}, \bibinfo {author} {\bibfnamefont
  {F.}~\bibnamefont {Recchia}}, \bibinfo {author} {\bibfnamefont
  {C.}~\bibnamefont {Walz}}, \bibinfo {author} {\bibfnamefont {D.}~\bibnamefont
  {Weisshaar}}, \ and\ \bibinfo {author} {\bibfnamefont {S.~J.}\ \bibnamefont
  {Williams}},\ }\href@noop {} {\bibfield  {journal} {\bibinfo  {journal}
  {Phys. Rev. C}\ }\textbf {\bibinfo {volume} {90}},\ \bibinfo {pages} {034301}
  (\bibinfo {year} {2014})}\BibitemShut {NoStop}%
\bibitem [{\citenamefont {Gade}\ \emph {et~al.}(2016)\citenamefont {Gade},
  \citenamefont {Tostevin}, \citenamefont {Bader}, \citenamefont {Baugher},
  \citenamefont {Bazin}, \citenamefont {Berryman}, \citenamefont {Brown},
  \citenamefont {Diget}, \citenamefont {Glasmacher}, \citenamefont {Hartley},
  \citenamefont {Lunderberg}, \citenamefont {Stroberg}, \citenamefont
  {Recchia}, \citenamefont {Ratkiewicz}, \citenamefont {Weisshaar},\ and\
  \citenamefont {Wimmer}}]{Gad16a}%
  \BibitemOpen
  \bibfield  {author} {\bibinfo {author} {\bibfnamefont {A.}~\bibnamefont
  {Gade}}, \bibinfo {author} {\bibfnamefont {J.~A.}\ \bibnamefont {Tostevin}},
  \bibinfo {author} {\bibfnamefont {V.}~\bibnamefont {Bader}}, \bibinfo
  {author} {\bibfnamefont {T.}~\bibnamefont {Baugher}}, \bibinfo {author}
  {\bibfnamefont {D.}~\bibnamefont {Bazin}}, \bibinfo {author} {\bibfnamefont
  {J.~S.}\ \bibnamefont {Berryman}}, \bibinfo {author} {\bibfnamefont {B.~A.}\
  \bibnamefont {Brown}}, \bibinfo {author} {\bibfnamefont {C.~A.}\ \bibnamefont
  {Diget}}, \bibinfo {author} {\bibfnamefont {T.}~\bibnamefont {Glasmacher}},
  \bibinfo {author} {\bibfnamefont {D.~J.}\ \bibnamefont {Hartley}}, \bibinfo
  {author} {\bibfnamefont {E.}~\bibnamefont {Lunderberg}}, \bibinfo {author}
  {\bibfnamefont {S.~R.}\ \bibnamefont {Stroberg}}, \bibinfo {author}
  {\bibfnamefont {F.}~\bibnamefont {Recchia}}, \bibinfo {author} {\bibfnamefont
  {A.}~\bibnamefont {Ratkiewicz}}, \bibinfo {author} {\bibfnamefont
  {D.}~\bibnamefont {Weisshaar}}, \ and\ \bibinfo {author} {\bibfnamefont
  {K.}~\bibnamefont {Wimmer}},\ }\href {\doibase 10.1103/PhysRevC.93.054315}
  {\bibfield  {journal} {\bibinfo  {journal} {Phys. Rev. C}\ }\textbf {\bibinfo
  {volume} {93}},\ \bibinfo {pages} {054315} (\bibinfo {year}
  {2016})}\BibitemShut {NoStop}%
\bibitem [{\citenamefont {Tostevin}(2001)}]{Tos01}%
  \BibitemOpen
  \bibfield  {author} {\bibinfo {author} {\bibfnamefont {J.~A.}\ \bibnamefont
  {Tostevin}},\ }\href@noop {} {\bibfield  {journal} {\bibinfo  {journal}
  {Nucl. Phys. A}\ }\textbf {\bibinfo {volume} {682}},\ \bibinfo {pages} {320}
  (\bibinfo {year} {2001})}\BibitemShut {NoStop}%
\bibitem [{\citenamefont {Honma}\ \emph {et~al.}(2002)\citenamefont {Honma},
  \citenamefont {Otsuka}, \citenamefont {Brown},\ and\ \citenamefont
  {Mizusaki}}]{Hon02}%
  \BibitemOpen
  \bibfield  {author} {\bibinfo {author} {\bibfnamefont {M.}~\bibnamefont
  {Honma}}, \bibinfo {author} {\bibfnamefont {T.}~\bibnamefont {Otsuka}},
  \bibinfo {author} {\bibfnamefont {B.~A.}\ \bibnamefont {Brown}}, \ and\
  \bibinfo {author} {\bibfnamefont {T.}~\bibnamefont {Mizusaki}},\ }\href@noop
  {} {\bibfield  {journal} {\bibinfo  {journal} {Phys. Rev. C}\ }\textbf
  {\bibinfo {volume} {65}},\ \bibinfo {pages} {061301} (\bibinfo {year}
  {2002})}\BibitemShut {NoStop}%
\bibitem [{\citenamefont {Honma}\ \emph {et~al.}(2004)\citenamefont {Honma},
  \citenamefont {Otsuka}, \citenamefont {Brown},\ and\ \citenamefont
  {Mizusaki}}]{Hon04}%
  \BibitemOpen
  \bibfield  {author} {\bibinfo {author} {\bibfnamefont {M.}~\bibnamefont
  {Honma}}, \bibinfo {author} {\bibfnamefont {T.}~\bibnamefont {Otsuka}},
  \bibinfo {author} {\bibfnamefont {B.~A.}\ \bibnamefont {Brown}}, \ and\
  \bibinfo {author} {\bibfnamefont {T.}~\bibnamefont {Mizusaki}},\ }\href@noop
  {} {\bibfield  {journal} {\bibinfo  {journal} {Phys. Rev. C}\ }\textbf
  {\bibinfo {volume} {69}},\ \bibinfo {pages} {034335} (\bibinfo {year}
  {2004})}\BibitemShut {NoStop}%
\bibitem [{\citenamefont {Honma}\ \emph {et~al.}(2005)\citenamefont {Honma},
  \citenamefont {Otsuka}, \citenamefont {Brown},\ and\ \citenamefont
  {Mizusaki}}]{Hon05}%
  \BibitemOpen
  \bibfield  {author} {\bibinfo {author} {\bibfnamefont {M.}~\bibnamefont
  {Honma}}, \bibinfo {author} {\bibfnamefont {T.}~\bibnamefont {Otsuka}},
  \bibinfo {author} {\bibfnamefont {B.~A.}\ \bibnamefont {Brown}}, \ and\
  \bibinfo {author} {\bibfnamefont {T.}~\bibnamefont {Mizusaki}},\ }\href@noop
  {} {\bibfield  {journal} {\bibinfo  {journal} {Eur. Phys. Journal A}\
  }\textbf {\bibinfo {volume} {25}},\ \bibinfo {pages} {499} (\bibinfo {year}
  {2005})}\BibitemShut {NoStop}%
\bibitem [{\citenamefont {Ormand}\ and\ \citenamefont {Brown}(1989)}]{Orm89}%
  \BibitemOpen
  \bibfield  {author} {\bibinfo {author} {\bibfnamefont {W.}~\bibnamefont
  {Ormand}}\ and\ \bibinfo {author} {\bibfnamefont {B.~A.}\ \bibnamefont
  {Brown}},\ }\href@noop {} {\bibfield  {journal} {\bibinfo  {journal} {Nucl.
  Phys. A}\ }\textbf {\bibinfo {volume} {491}},\ \bibinfo {pages} {1} (\bibinfo
  {year} {1989})}\BibitemShut {NoStop}%
\bibitem [{\citenamefont {Brown}(2014)}]{NuX}%
  \BibitemOpen
  \bibfield  {author} {\bibinfo {author} {\bibfnamefont {B.~A.}\ \bibnamefont
  {Brown}},\ }\href@noop {} {\bibfield  {journal} {\bibinfo  {journal} {Nuclear
  Data Sheets}\ }\textbf {\bibinfo {volume} {120}},\ \bibinfo {pages} {115}
  (\bibinfo {year} {2014})}\BibitemShut {NoStop}%
\bibitem [{\citenamefont {Wang}\ \emph {et~al.}(2017)\citenamefont {Wang},
  \citenamefont {Audi}, \citenamefont {Kondev}, \citenamefont {Huang},
  \citenamefont {Naimi},\ and\ \citenamefont {Xu}}]{Wang17a}%
  \BibitemOpen
  \bibfield  {author} {\bibinfo {author} {\bibfnamefont {M.}~\bibnamefont
  {Wang}}, \bibinfo {author} {\bibfnamefont {G.}~\bibnamefont {Audi}}, \bibinfo
  {author} {\bibfnamefont {F.~G.}\ \bibnamefont {Kondev}}, \bibinfo {author}
  {\bibfnamefont {W.}~\bibnamefont {Huang}}, \bibinfo {author} {\bibfnamefont
  {S.}~\bibnamefont {Naimi}}, \ and\ \bibinfo {author} {\bibfnamefont
  {X.}~\bibnamefont {Xu}},\ }\href {\doibase 10.1088/1674-1137/41/3/030003}
  {\bibfield  {journal} {\bibinfo  {journal} {Chinese Physics C}\ }\textbf
  {\bibinfo {volume} {41}},\ \bibinfo {pages} {030003} (\bibinfo {year}
  {2017})}\BibitemShut {NoStop}%
\bibitem [{\citenamefont {Bentley}\ and\ \citenamefont {Lenzi}(2007)}]{Ben07a}%
  \BibitemOpen
  \bibfield  {author} {\bibinfo {author} {\bibfnamefont {M.}~\bibnamefont
  {Bentley}}\ and\ \bibinfo {author} {\bibfnamefont {S.}~\bibnamefont
  {Lenzi}},\ }\href {\doibase https://doi.org/10.1016/j.ppnp.2006.10.001}
  {\bibfield  {journal} {\bibinfo  {journal} {Prog. Part. Nucl. Phys.}\
  }\textbf {\bibinfo {volume} {59}},\ \bibinfo {pages} {497} (\bibinfo {year}
  {2007})}\BibitemShut {NoStop}%
\bibitem [{Wig()}]{Wig57}%
  \BibitemOpen
  \href@noop {} {}\bibinfo {howpublished} {E.P. Wigner, in {\it Proceedings of
  the Robert A. Welch Foundation Conference on Chemical Research}, edited by
  W.O. Milligan (Welch Foundation, Houston, 1957), Vol.1}\BibitemShut {NoStop}%
\bibitem [{\citenamefont {Bentley}\ \emph {et~al.}(2015)\citenamefont
  {Bentley}, \citenamefont {Lenzi}, \citenamefont {Simpson},\ and\
  \citenamefont {Diget}}]{Ben15}%
  \BibitemOpen
  \bibfield  {author} {\bibinfo {author} {\bibfnamefont {M.~A.}\ \bibnamefont
  {Bentley}}, \bibinfo {author} {\bibfnamefont {S.~M.}\ \bibnamefont {Lenzi}},
  \bibinfo {author} {\bibfnamefont {S.~A.}\ \bibnamefont {Simpson}}, \ and\
  \bibinfo {author} {\bibfnamefont {C.~A.}\ \bibnamefont {Diget}},\ }\href@noop
  {} {\bibfield  {journal} {\bibinfo  {journal} {Phys. Rev. C}\ }\textbf
  {\bibinfo {volume} {92}},\ \bibinfo {pages} {024310} (\bibinfo {year}
  {2015})}\BibitemShut {NoStop}%
\bibitem [{\citenamefont {Spieker}\ \emph {et~al.}(2019)\citenamefont
  {Spieker}, \citenamefont {Gade}, \citenamefont {Weisshaar}, \citenamefont
  {Brown}, \citenamefont {Tostevin}, \citenamefont {Longfellow}, \citenamefont
  {Adrich}, \citenamefont {Bazin}, \citenamefont {Bentley}, \citenamefont
  {Brown}, \citenamefont {Campbell}, \citenamefont {Diget}, \citenamefont
  {Elman}, \citenamefont {Glasmacher}, \citenamefont {Hill}, \citenamefont
  {Pritychenko}, \citenamefont {Ratkiewicz},\ and\ \citenamefont
  {Rhodes}}]{Spi19a}%
  \BibitemOpen
  \bibfield  {author} {\bibinfo {author} {\bibfnamefont {M.}~\bibnamefont
  {Spieker}}, \bibinfo {author} {\bibfnamefont {A.}~\bibnamefont {Gade}},
  \bibinfo {author} {\bibfnamefont {D.}~\bibnamefont {Weisshaar}}, \bibinfo
  {author} {\bibfnamefont {B.~A.}\ \bibnamefont {Brown}}, \bibinfo {author}
  {\bibfnamefont {J.~A.}\ \bibnamefont {Tostevin}}, \bibinfo {author}
  {\bibfnamefont {B.}~\bibnamefont {Longfellow}}, \bibinfo {author}
  {\bibfnamefont {P.}~\bibnamefont {Adrich}}, \bibinfo {author} {\bibfnamefont
  {D.}~\bibnamefont {Bazin}}, \bibinfo {author} {\bibfnamefont {M.~A.}\
  \bibnamefont {Bentley}}, \bibinfo {author} {\bibfnamefont {J.~R.}\
  \bibnamefont {Brown}}, \bibinfo {author} {\bibfnamefont {C.~M.}\ \bibnamefont
  {Campbell}}, \bibinfo {author} {\bibfnamefont {C.~A.}\ \bibnamefont {Diget}},
  \bibinfo {author} {\bibfnamefont {B.}~\bibnamefont {Elman}}, \bibinfo
  {author} {\bibfnamefont {T.}~\bibnamefont {Glasmacher}}, \bibinfo {author}
  {\bibfnamefont {M.}~\bibnamefont {Hill}}, \bibinfo {author} {\bibfnamefont
  {B.}~\bibnamefont {Pritychenko}}, \bibinfo {author} {\bibfnamefont
  {A.}~\bibnamefont {Ratkiewicz}}, \ and\ \bibinfo {author} {\bibfnamefont
  {D.}~\bibnamefont {Rhodes}},\ }\href {\doibase 10.1103/PhysRevC.99.051304}
  {\bibfield  {journal} {\bibinfo  {journal} {Phys. Rev. C}\ }\textbf {\bibinfo
  {volume} {99}},\ \bibinfo {pages} {051304} (\bibinfo {year}
  {2019})}\BibitemShut {NoStop}%
\end{thebibliography}%

\end{document}